\documentclass[]{beilstein}

\usepackage{xspace}

\usepackage{amsmath,amssymb}
\usepackage[]{color}
\usepackage[]{float}
\usepackage[version=3]{mhchem}
\usepackage[]{graphicx}

\begin{document}


\title{Simulation of electron transport in electron-beam-induced deposition of
nanostructures}
\author*{Francesc Salvat-Pujol}{salvat-pujol@itp.uni-frankfurt.de}
\affiliation{Institut f\"ur Theoretische Physik, Goethe-Universit\"at Frankfurt, Max-von-Laue-Stra{\ss}e 1, 60438 Frankfurt am Main, Germany}
\author[1]{Harald O.\ Jeschke}
\author[1]{Roser Valent\'{i}}
\maketitle
\begin{abstract}
We present a numerical investigation of energy and charge distributions
during electron-beam-induced growth of W nanostructures on SiO$_2$
substrates using Monte Carlo simulation of electron transport. This
study gives a quantitative insight into the deposition of energy and
charge in the substrate and in already existing metallic nanostructures
in the presence of the electron beam. We analyze electron trajectories,
inelastic mean free paths, and distribution of backscattered electrons
in different deposit compositions and depths. We find that, while in the
early stages of the nanostructure growth a significant fraction of
electron trajectories still interact with the substrate, as the
nanostructure becomes thicker the transport takes place almost
exclusively in the nanostructure. In particular, a larger deposit
density leads to enhanced electron backscattering. This work shows how
mesoscopic radiation-transport techniques can contribute to a model
which addresses the multi-scale nature of the electron-beam-induced
deposition (EBID) process. Furthermore, similar simulations can aid in
understanding the role played by backscattered electrons and emitted
secondary electrons in the change of structural properties of
nanostructured materials during post-growth electron-beam treatments.
\end{abstract}


\section{Introduction}

Electron-beam-induced deposition
(EBID)~\cite{koops1993,randolph2006,utke2008}
is a method suitable for template-free fabrication of nanostructures.
Molecules of a precursor gas are injected into a high- or
ultra-high-vacuum chamber and are dissociated by a 1-50 keV focussed
electron beam into a volatile fragment, which is evacuated by the vacuum
system, and a non-volatile fragment, which is progressively adsorbed on
a substrate, thus leading to the growth of a nanostructure at the focus
of the beam. In general, the obtained deposits exhibit a granular
structure consisting of nanometer-sized metal crystallites which are
embedded in an insulating matrix. 

There are three main interactions that determine the growth of
nanostructures in the EBID process: (1) substrate-precursor interaction,
(2) electron-substrate interaction and (3) electron-precursor
interaction. In this work we concentrate on the electron-substrate
interaction and our results have some implications for
electron-precursor interaction. Existing theories for the EBID process
\cite{huth2012}
consist mainly of equations for the deposition rate that can either be
solved analytically under simplifying assumptions or in a more general
form using Monte Carlo simulations. However, there is no theory which
addresses the multi-scale nature of the EBID process, including
microscopic and mesoscopic length and time scales, from ultra-fast
(non-equilibrium processes occurring in femtoseconds) to relatively slow
(growth and relaxation processes requiring nanoseconds or even
microseconds).

In this work we focus on the mesoscopic length scale and present a
detailed numerical study of the distribution of energy and charge
originating from EBID conditions. The study is not only relevant for
EBID, but also as a first step for understanding aspects of other
experimental techniques including, \textit{e.\ g.}, the effect of
backscattered electrons in changing structural properties in direct and
oxygen-assisted electron-beam post-growth nanostructure treatments
\cite{porrati2011,mehendale2013}.  We
consider various geometric settings as well as materials relevant to
EBID nanostructure growth. For our simulations we use the Monte Carlo
code for radiation transport \textsc{Penelope}~\cite{salvat}, where a
statistical set of particle trajectories are sampled in homogeneous
materials. In this context, we provide an overview of the aspects of
EBID nanostructure growth that can be studied in detail from a
mesoscopic point of view using well-established radiation-transport
simulation techniques for amorphous media \cite{joy,lin2005}. Recently,
practical Monte Carlo simulations of EBID-nanostructure growth have been
reported \cite{liu2006,silvis-cividjian2002,smith2007,smith2008} on the
basis of simplified transport models based, \textit{e.\ g.}, on the
Rutherford cross section or on a plural-scattering scheme whereby the
inelastic scattering of electrons in solids is treated in an average
fashion using the continuous slowing-down approximation (in this
approximation only the energy loss per unit path length is respected:
energy fluctuations are not captured). In the present work we sample
inelastic interactions in detail, \textit{i.e.}, on a per-interaction
basis without employing a condensed simulation scheme, and we restrict
our considerations to the interaction of the primary electrons with the
substrate and the nanostructure at different stages of its growth.

The precursor gas we shall consider throughout this study is tungsten
hexacarbonyl, W(CO)$_6$, and the corresponding deposits W$_x$C$_y$O$_z$,
{\it i.e.} amorphous tungsten oxycarbides with varying carbon and oxygen
contents.  W(CO)$_6$ belongs to the class of organometallic compounds
that are well established for the EBID
process~\cite{hoyle1993,hoyle1996,huth2009}.
It has been studied in detail  by mass
spectrometry~\cite{beranova1994,cooks1990,michels1980} and photoelectron or
photoionization spectroscopy~\cite{cooper1987,hubbard1982,qi1997}, which yield
appearance energies of ionic fragments as well as approximate internal
energy distributions after electron ionization.  The main advantage of
using this precursor gas is that the tungsten metal content in the
deposits can be widely varied so as to cover a wide range of electronic
properties, from insulating to metallic \cite{huth2009,muthukumar2012}. Our aim
is to determine a spatially resolved picture of the growth conditions
created by the electron beam within and above a SiO$_2$ substrate as
well as within and above W$_x$C$_y$O$_z$ deposits of various
thicknesses.

\section{Description of the simulation}

\begin{figure}
  \centering
  \includegraphics[width=0.6\textwidth]{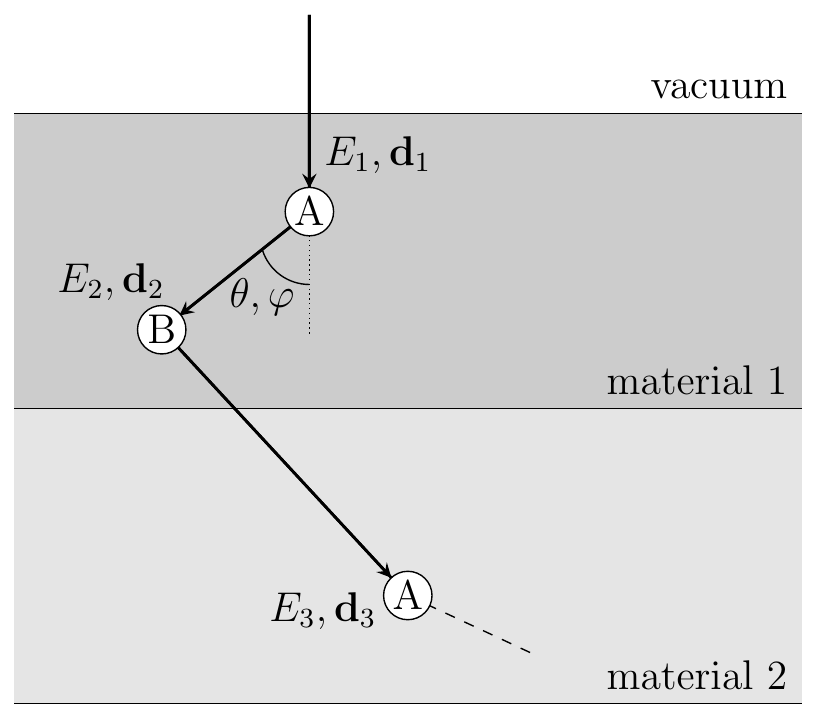}
  \caption{Schematic representation of a random trajectory generated by
    \textsc{Penelope}~\protect\cite{salvat}. The trajectory is determined by
    path lengths $s$ that determine the position ${\bf r}_n$ of the
    next scattering event, by the type of event, and by energies $E_n$
    and directions ${\bf d}_n$ after the event.}
 \label{fig:randomtrajectory}
\end{figure}

The Monte Carlo method for the simulation of radiation transport is a
numerical means of solving the Boltzmann transport equation in an
arbitrary geometry. The computer code system \textsc{Penelope} yields
trajectories of primary and secondary particles according to state-of-the-art
interaction cross sections on sample geometries constructed by
positioning a set of well-defined homogeneous bodies in space.  Random
trajectories are generated as follows~\cite{salvat}: particles are
characterized by their position vector ${\bf r}=(x,y,z)$, energy $E$ and
a direction-of-flight unit vector ${\bf d}=(u,v,w)$, where $u$, $v$, and
$w$ are the direction cosines. A particle trajectory is represented as a
series of states $({\bf r}_n, E_n, {\bf d}_n)$ where $n$ labels the
scattering event at ${\bf r}_n$ leading to energy $E_n$ and direction
${\bf d}_n$ (see Figure~\ref{fig:randomtrajectory}).  Several random
variables are sampled from their respective probability distribution
functions. The length of the free path to the next collision, $s$, is
sampled from an exponential distribution with total mean free path
$\lambda_T$ using a random number $\xi$ uniformly distributed in the
interval $(0,1)$,
\begin{equation}
s= -\lambda_{\rm T} \ln \xi.
\end{equation}
The interaction type at the new position is sampled as follows. Let us
consider interactions of type A and B, with respective total cross
sections $\sigma_{\rm A}$ and $\sigma_{\rm B}$. Interactions of type A
and B are sampled with probabilities
\begin{equation}
p_{\rm A}=\frac{\sigma_{\rm A}}{\sigma_{\rm T}}\,,\qquad p_{\rm B}=\frac{\sigma_{\rm B}}{\sigma_{\rm T}}\,,
\end{equation}
respectively, where $\sigma_{\rm T}=\sigma_{\rm
A}+\sigma_{\rm B}$ is the total interaction cross section. The polar
scattering angle $\theta$ and the energy loss $W$ are sampled from a
distribution with azimuthal symmetry,
\begin{equation}
p_{{\rm A},{\rm B}}(E;\theta,W) = \frac{2\pi\sin \theta}{\sigma_{{\rm
A},{\rm B}}(E)}\frac{d^2 \sigma_{{\rm A},{\rm B}}(E;\theta,W)}{d\Omega
dW}.
\end{equation}
Finally, the azimuthal scattering angle is sampled from a uniform
random number $\xi$ as $\phi=2 \pi \xi$.

The \textsc{Penelope} code \cite{salvat} uses a relatively
sophisticated interaction model, devised for energies above a few
hundred eV. Differential cross sections for elastic scattering were
calculated with the state-of-the-art relativistic partial-wave
calculation code \textsc{Elsepa} \cite{salvat2005}. Inelastic interactions
are described by means of the plane-wave Born approximation using a
schematized generalized-oscillator-strength model, fitted to reproduce
at high energies the stopping power obtained from the asymptotic Bethe
formula. 

\begin{figure}
  \centering
  \includegraphics[width=0.45\textwidth]{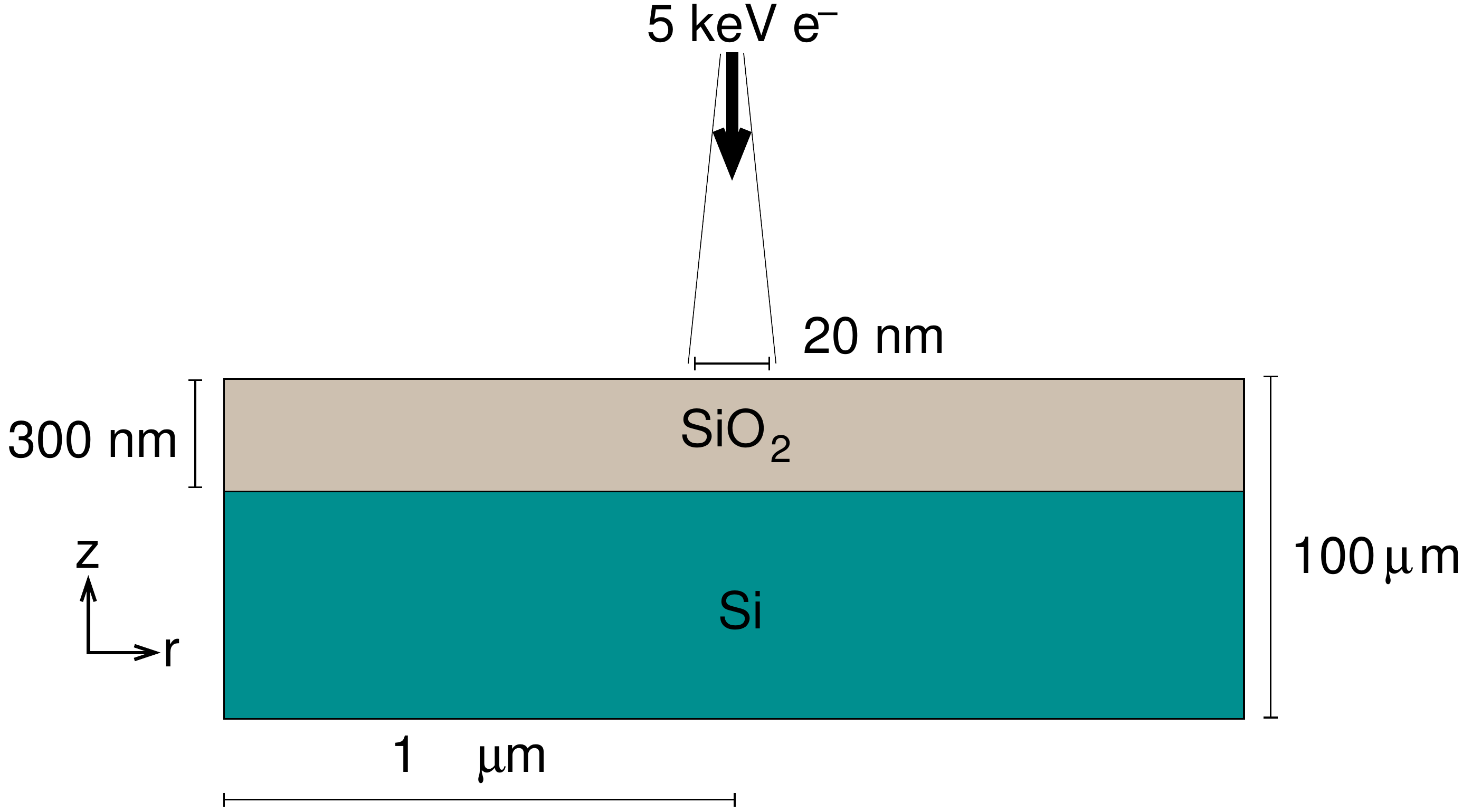}
  \hspace{0.05\textwidth}
  \includegraphics[width=0.45\textwidth]{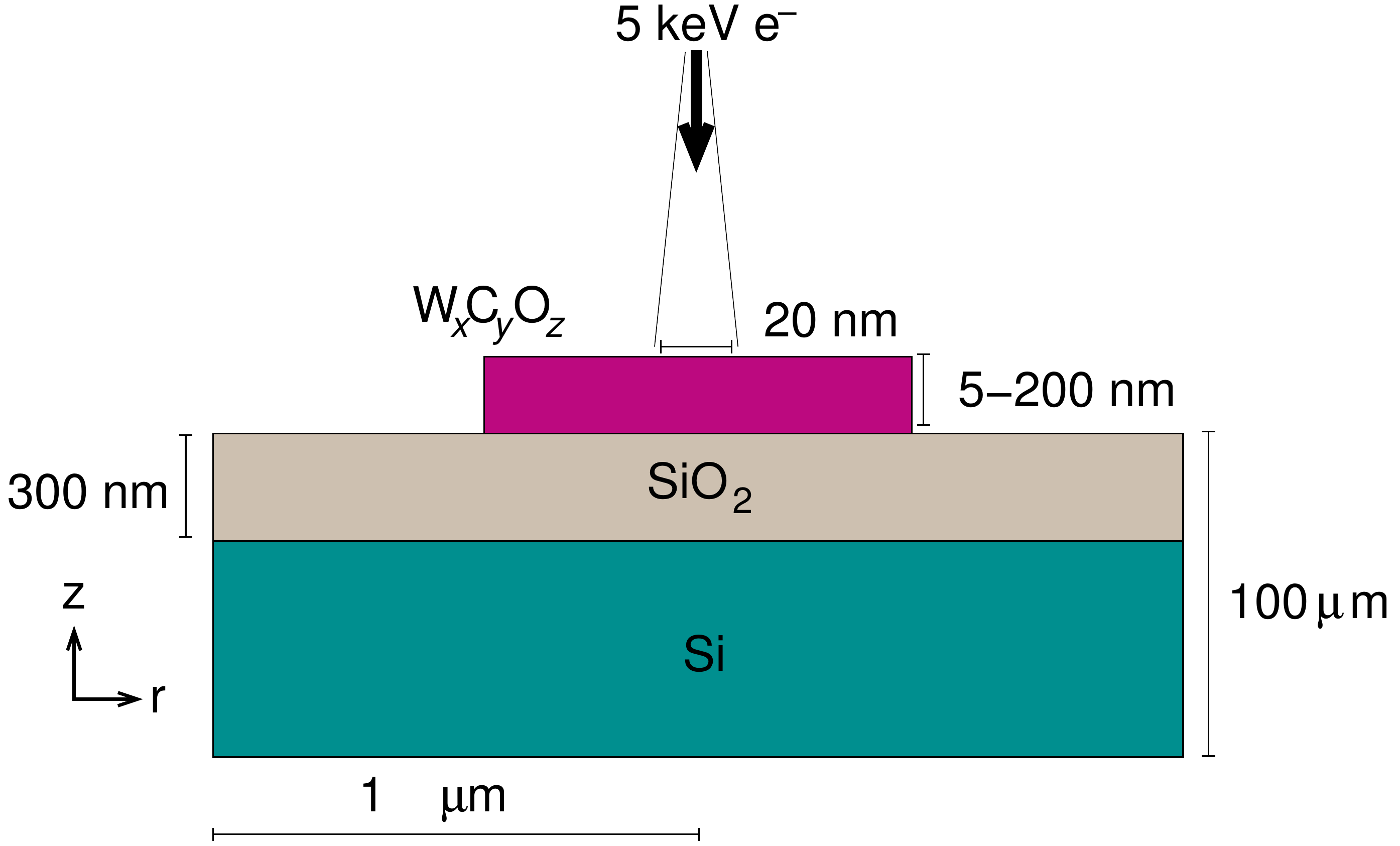}
  \caption{Cylindrical sample geometries used in the simulations. Left:
    A 300 nm thick amorphous SiO$_2$ substrate is placed above a Si
    wafer and irradiated with 5 keV electrons. Right: A W$_x$C$_y$O$_z$
    deposit of thickness from 5~nm to 200~nm and of density and
    composition as given in Table \ref{tab:dens} is placed above a 300
    nm amorphous SiO$_2$ layer, which in turn is placed on top of a Si
    wafer. A conical electron beam with a spot size of 20 nm on the
    sample is used, the point source being located 1 cm in vacuum above
    the center of the sample; the corresponding beam aperture is
    5.73$\times10^{-5}$ degrees. A radius of 100 nm is chosen for the
    deposit.
  }
 \label{fig:geometry}
\end{figure}

In our study it is convenient to reduce the problem to two spatial
dimensions by assuming a geometry with cylindrical symmetry. We perform
studies for two classes of sample geometries: (a)  a 300 nm thick layer
of amorphous SiO$_2$ with density 2.32 g/cm$^{3}$ is placed on top of a
Si wafer of density 2.33 g/cm$^{3}$ in order to study the initial
conditions of the EBID growth process. We refer to this sample geometry
briefly as the ``substrate''.
(b) Structures corresponding to intermediate EBID deposits are
constructed in order to study the conditions for further growth in the
EBID process, where deposited layers of different thicknesses (from 5
nm to 200 nm) are placed on top of the substrate surface, with its
density and composition set in accordance with 6 different
experimentally realized EBID structures~\cite{huth2009}. While the
composition in terms of atomic percent is taken from
Ref.~\cite{huth2009}, the densities were determined in
Ref.~\cite{muthukumar2012} by predicting approximate crystal structures
at these given compositions using evolutionary-algorithm-based crystal
structure prediction. The composition and densities of the deposits
are given in Table~\ref{tab:dens}.
In both cases, a beam of 5-keV electrons impinges normally on the
surface, with a spot size of 20 nm diameter. In practice, the electron
beam is rastered on the substrate, so that the extension of the
deposited nanostructure can be larger than the electron-beam spot size.
Thus, a radius of 100~nm has been taken for the deposited nanostructure.
The linear range of 5 keV electrons in Si and W is about 0.4 $\mu$m and
0.1 $\mu$m, respectively. Thus, in order to ensure that virtually no
electrons leave the simulation geometry through the lateral bounds
(direction perpendicular to the incoming direction), a
cylinder radius of 1$\mu$m has been set.
See Figure~\ref{fig:geometry} for an illustration.

\begin{table}
  \begin{center}
    \begin{tabular}{|ccccc|}
    \hline
     Composition & Density (g/cm$^3$) & W (at \%) & C (at \%) & O (at \%)\\ 
     approximant&&&&\\ \hline
     WC$_{2.5}$O         &  7.9  &  22.6 & 56.0 & 21.4\\ 
     WC$_{3.33}$O$_{0.67}$ &  8.7  &  19.0 & 67.1 & 13.8\\ 
     WC$_{1.4}$O$_{0.8}$  &  9.1  &  31.8 & 44.4 & 23.8\\ 
     WCO$_{0.71}$        &  10.0  &  36.9 & 35.6 & 27.5\\ 
     WC$_{1.33}$O$_{0.67}$ &  10.4  &  34.0 & 44.3 & 21.7\\ 
     WC$_{1.75}$O$_{0.75}$ &  10.6  &  27.5 & 50.4 & 22.1\\ \hline 
    \end{tabular}
    \caption{Composition of the six amorphous tungsten oxycarbide
      deposits considered in this study, following Ref.~\cite{huth2009}
      and given in terms of atomic percent (at \%).  They are sorted
      by increasing density which was determined in
      Ref.~\cite{muthukumar2012} (see text).}
    \label{tab:dens}
  \end{center}
\end{table}

Material cross sections in our calculations are approximated as an
averaged weighted sum of the atomic cross sections corresponding to a
given composition (incoherent sum of scattered intensities), thus
neglecting chemical binding effects.  Energetic electrons can scatter
either elastically, where the quantum state of the scatterer remains
unaltered and the direction of the projectile changes, or inelastically,
where electronic excitations or ionizations take place through the
different energy and momentum transfer channels available. As the
electrons evolve through the medium, they lose energy in the course of
several inelastic interactions. The lost energy is either absorbed by
the medium through local excitations, which are allowed to relax through
the emission of photons, or through ionization of the sample, which
leads to the build-up of a localized positive charge in the material and
to new particles, thus leading to a ``shower'' of particles. If an
electron crosses a boundary into an adjacent material, its trajectory history
is stopped at the other side of the interface and restarted with the new
material transport properties. This can be done any time, since electron
trajectories are modelled as Markov processes (the future of the
trajectory is dependent only on the present state, and not on the past).
The trajectory history of an electron is stopped when its energy drops below
50 eV: the electron is then considered absorbed by the medium,
contributing to the build-up of a localized negative charge in the
material. We choose an absorption energy of 50 eV because we are
neglecting binding effects in the material, and furthermore, elastic and
inelastic cross sections derived from atomistic models carry large
uncertainties already for energies below a few hundred eV. The same
absorption energy is used for those secondary electrons generated in the
shower. Finally, to obtain our simulated results, we have sampled 10$^8$
trajectories.

\section{Results}

\begin{figure}
  \centering
  \includegraphics[scale=0.25]{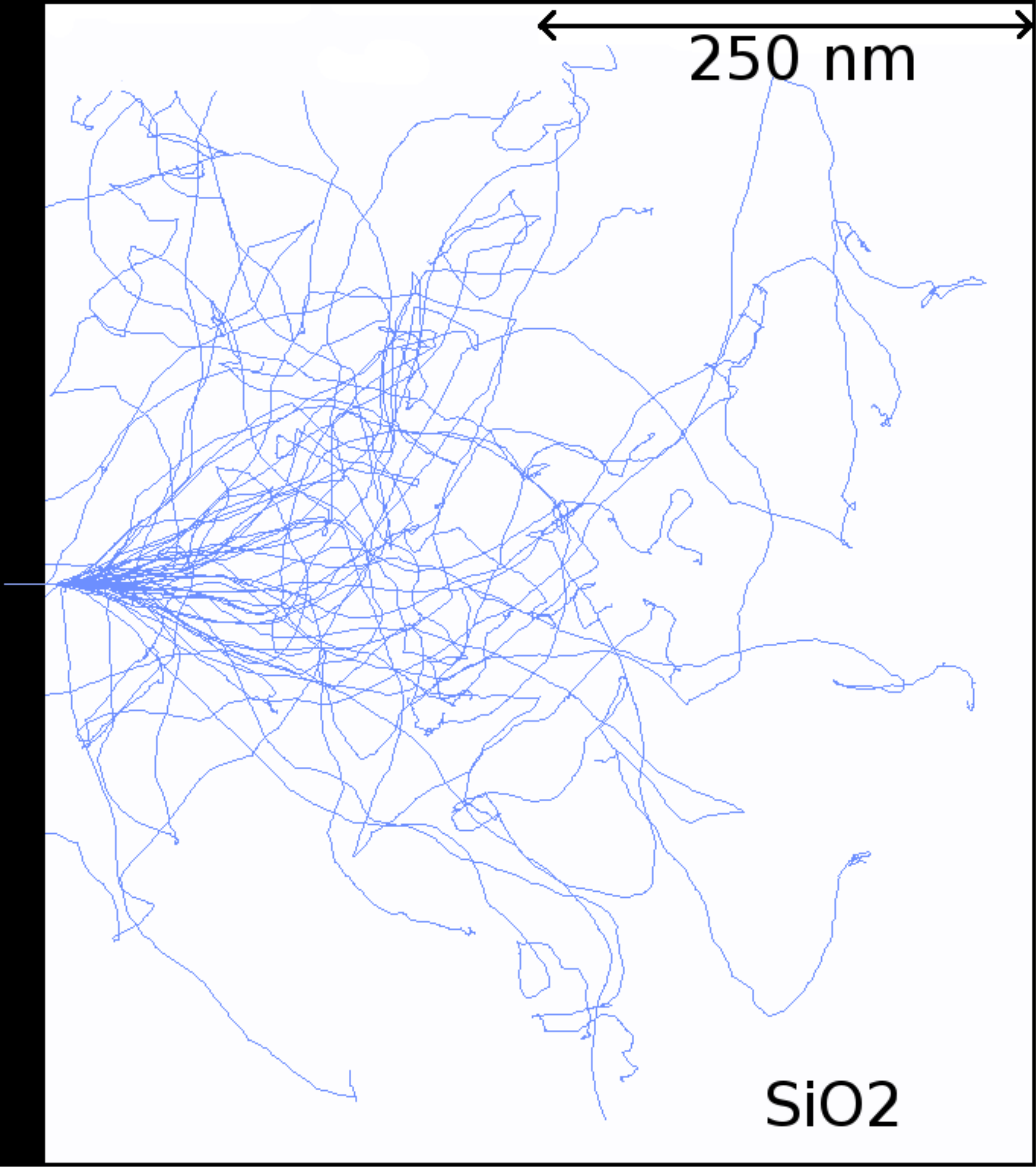}
  \hspace{1cm}%
  \includegraphics[scale=0.25]{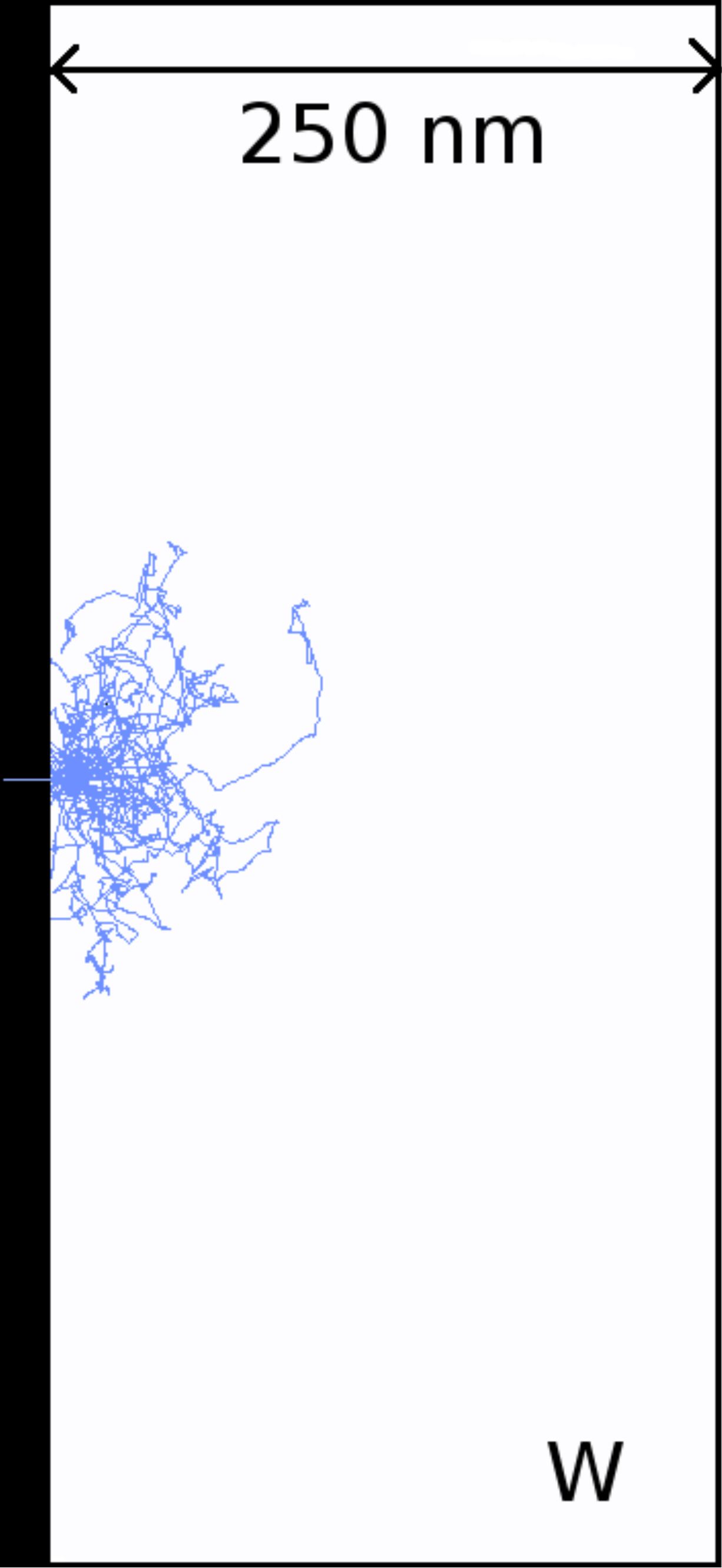}%
  \caption{Snapshot of 50 simulated electron trajectories in the SiO$_2$
  substrate (left-hand side) and in the nanostructure deposit material W
  (right-hand side, representative for material deposit). The width of
  the screenshot windows corresponds to 500~nm.}
  \label{fig:shower}
\end{figure}

\begin{figure}
  \centering
  \includegraphics[width=0.75\textwidth]{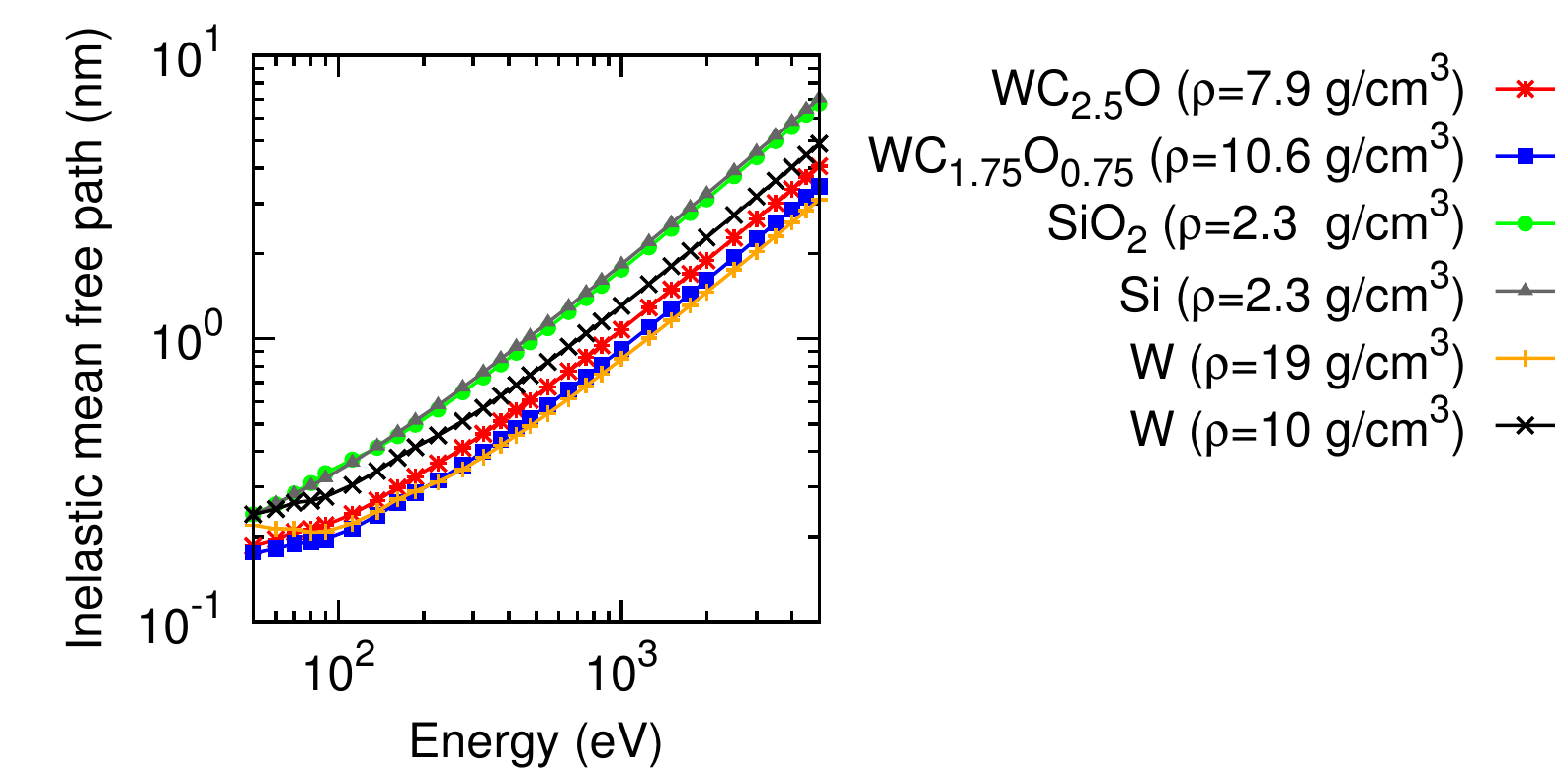}
  \caption{Inelastic mean free paths for the relevant
    materials in this work presented in the usual log-log scale. The
    tungsten oxycarbide compositions WC$_{2.5}$O and
    WC$_{1.75}$O$_{0.75}$ correspond to the lowest and highest density
    samples of Table~\ref{tab:dens}, respectively. The two curves with
    cross symbols show the variation due to a density change only.}
  \label{fig:imfp}
\end{figure}

To provide a first visual insight into the electron transport process in
the substrate and in the deposited nanostructure,
Figure~\ref{fig:shower} displays a simulated shower of 5-keV-electron
trajectories impinging normally on a 500 nm thick slab of SiO$_2$
(left-hand side, substrate material) and on a 500 nm thick slab of pure
W (right-hand side, deposit material), respectively.  We consider pure
tungsten as a representative material of the different deposits for
practical reasons. This choice is reasonable inasmuch as the average
distance between consecutive inelastic collisions [inelastic mean free
path (IMFP)] of electrons in W and in the considered nanostructure
materials are very similar in the energy window of interest (see
Figure~\ref{fig:imfp}). Note that in the SiO$_2$ substrate the beam is
completely attenuated at a depth of $\sim500$~nm, whereas in W this
depth is reduced to $\sim150$ nm.  Indeed, the IMFP of electrons in
SiO$_2$ is roughly a factor 2--4 larger than the IMFP of electrons in W
(or any of the 6 considered deposit materials), as shown on the
left-hand panel of Figure~\ref{fig:imfp}. Thus, we conclude that in the
early stages of the nanostructure growth (thicknesses much smaller than
\mbox{$\sim150$ nm}), the electron beam probes both the thin deposit and
the substrate. The energy and charge deposition processes are therefore
dictated by the transport characteristics of both the deposit and the
substrate. On the other hand, for nanostructure thicknesses exceeding
\mbox{$\sim150$ nm}, the deposition of energy and charge takes place
almost exclusively in the nanostructure, without affecting the
substrate. A similar analysis has been carried out in
Ref.~\cite{smith2008}. Experimentally, similar conclusions are drawn
from current measurements \cite{bret2003}.

\begin{figure}
  \centering
  \includegraphics[width=0.48\textwidth]{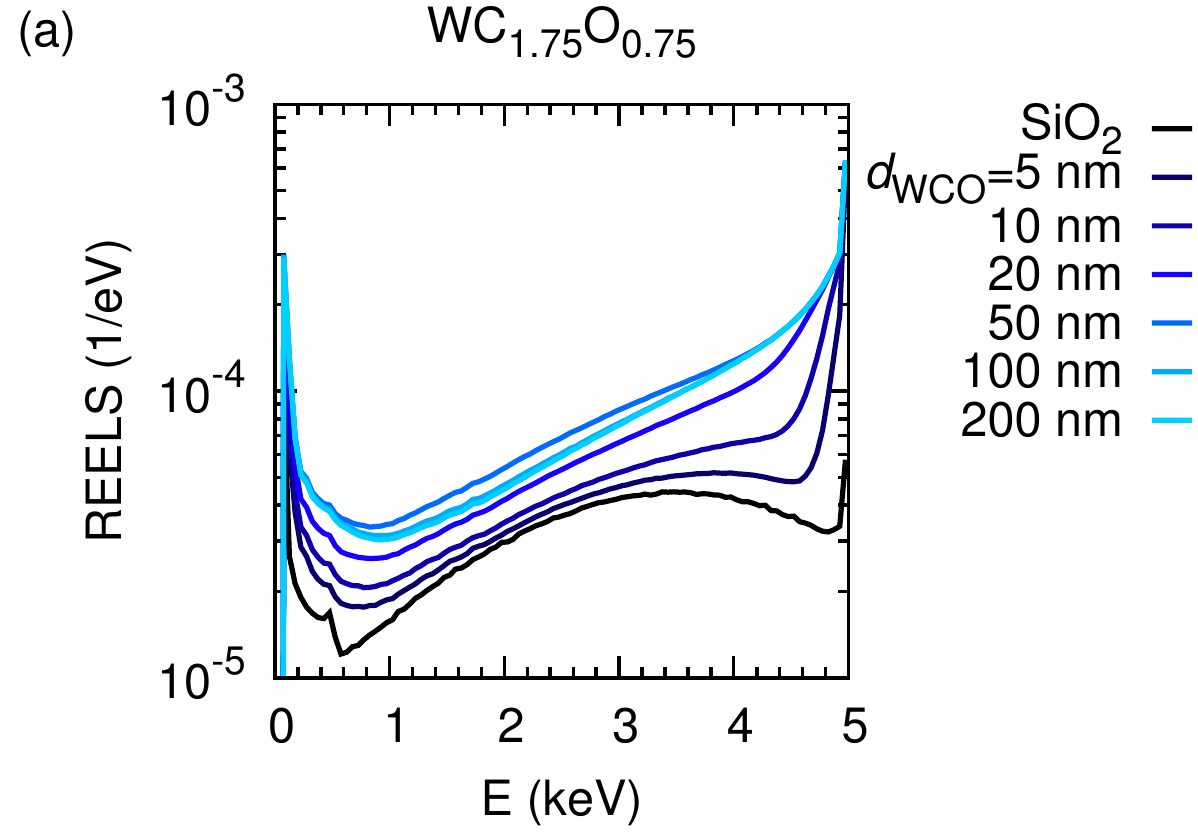}
  \hfill
  \includegraphics[width=0.48\textwidth]{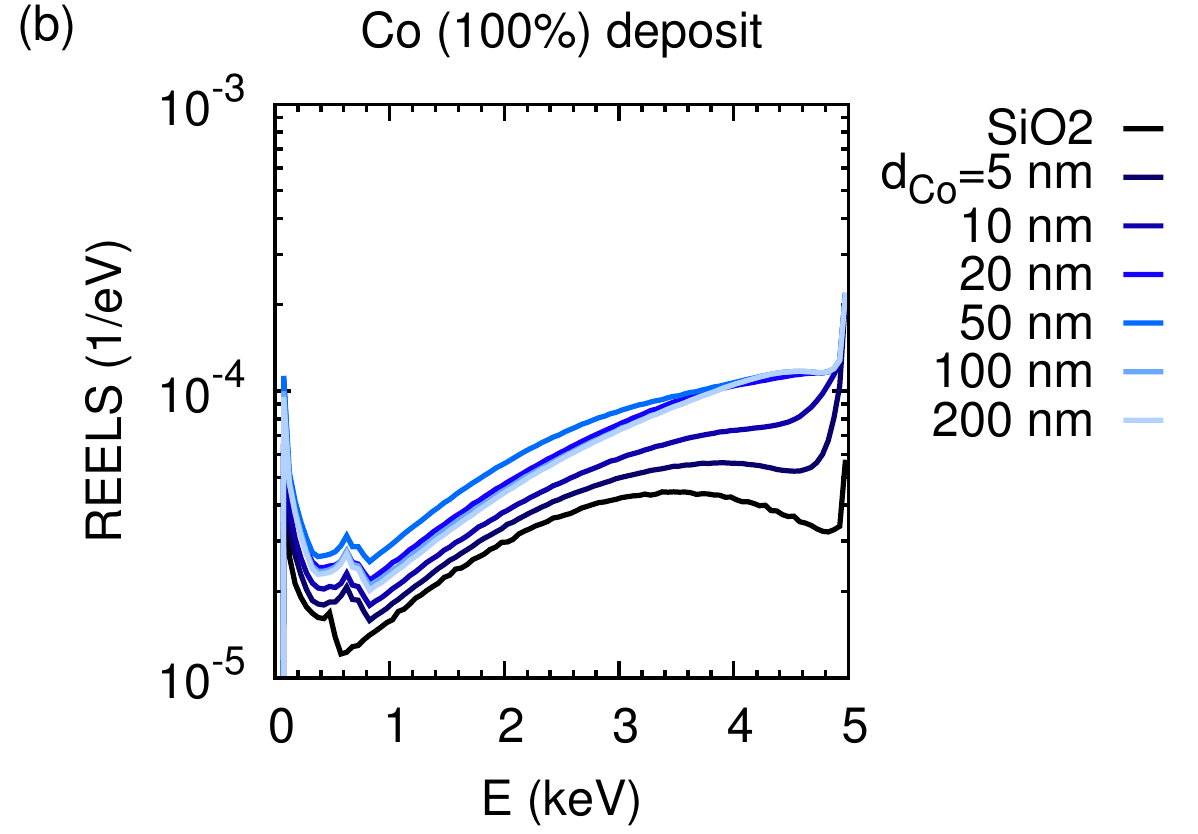}
  \caption{(a) Distribution of electrons backscattered and
    emitted into the vacuum from the substrate in absence of a deposit
    (black curve) and from the substrate with a deposit of thickness
    $d_\text{WCO}$ ranging from 5~nm to 200~nm (rest of curves)
    consisting of the material corresponding to the composition
    WC$_{1.75}$O$_{0.75}$ (see Table~\ref{tab:dens}). (b) Same as panel
    (a) for a deposit of pure Co. Notice that the ordinates are in a
  logarithmic scale, whereas the abscissas are in a linear scale. The
acronym REELS stands for reflection electron-energy-loss spectrum.}
  \label{fig:reels}
\end{figure}

Figure~\ref{fig:reels}(a) displays the energy
distribution of electrons backscattered and emitted per incoming
electron from the substrate (darkest curve) and from deposits of
increasing thicknesses $d_\text{WCO}$ on top of the substrate [dark blue
curve, $d_\text{WCO}=5$ nm, through light blue curve, $d_\text{WCO}=200$
nm]. Notice that for thin deposits the spectral features of the
substrate are merely smeared out, owing to the fact that only few
inelastic interactions take place in the thin deposit. For increasing
deposit thicknesses, the transport in the substrate plays an
increasingly marginal role. Thus, for thick deposits the spectral
features of the substrate vanish and the spectral features of the
deposit prevail. This explains the saturated behavior of the curves
corresponding to $d_\text{WCO}=100$ nm and $d_\text{WCO}=200$ nm, where
electrons are very unlikely to even reach the substrate, in accordance
with the discussion of Figure~\ref{fig:shower}. It is interesting to
note that the intensity in the energy distribution of backscattered
electrons increases with the sample thickness. Indeed, on the one hand
the elastic backscattering coefficient increases with the atomic number,
leading to the observed increase in the elastic peak at 5 keV [the
substrate consists of Si and O (atomic numbers $Z=14$ and $Z=8$,
respectively) whereas the deposit material contains W ($Z=74$)]. On the
other hand, the IMFP is inversely proportional to the material density,
so that a denser deposit on a comparatively light substrate implies an
increase in the number of energy losses per unit path length with
respect to those that would take place in the substrate alone. This
justifies the factor $\sim2$ between the curves corresponding to the
(thick) deposit and the substrate. Thus, under the assumption that the
presence of a large number of electrons (slow or fast) enhances the
dissociation rate of precursor gas molecules adsorbed on the substrate,
one can infer the following positive-feedback process: as the deposit
thickness grows, so does the number of backscattered and emitted
electrons, leading to an improvement in the dissociation rate and,
therefore, in the nanostructure deposition rate, leading to a reinforced
growth process. A more quantitative description of the change in
deposition rate would imply analyzing the separate contributions from
backscattered electrons, forward-scattered electrons, and secondary
electrons \cite{plank2012}. 

Two aspects of Figure~\ref{fig:reels}(a) should be emphasized. (1) In
order to further elucidate the dependence of the electron backscattering
probability with the atomic number of the deposit material, the
simulation was repeated replacing the deposit with Co, a comparatively
lighter material ($Z=27$). Figure~\ref{fig:reels}(b) displays the energy
distribution of backscattered electrons for different Co-nanodeposit
thicknesses, $d_\text{Co}$. Notice that the increase in the elastic-peak
intensity is roughly a factor 2 or 3 smaller than for the nanostructure
material, which is much heavier. (2) Notice that as the deposit becomes
thicker, the intensity of the curves increases monotonically, reaching
its maximum for a thickness of about 50 nm and then decreasing slightly
into its saturated value for thicknesses of 200 nm. The fact that
multiple elastic and inelastic interactions take place along the
trajectory makes it hard to give a detailed explanation of this effect.
Nevertheless, it can be argued that for thicknesses exceeding 50 nm, the
fraction of trajectories which reach the substrate becomes negligible
and, for thick enough deposits, this fraction approaches zero. Owing to
the fact that the mean free paths in the deposit are much shorter than
in the substrate, more energy losses take place per unit path length in
the deposit than in the substrate. This implies that the thicker the
deposit becomes, the larger is the number of electrons which leave the
sample after losing most of its energy. This explains, at least
qualitatively, the increase and eventual saturation in the low-energy
regime of the spectrum (contribution of electrons which leave the sample
after losing most of its energy and of emitted secondary electrons), as
well as the decrease in intensity in the energy range between 1 and 4
keV (for thick substrate electrons in this regime lose more energy and
therefore the spectral intensity shifts to lower energies). Close to the
elastic peak, variations with the thickness of the substrate between
50~nm and 200~nm are not visible, since the elastic backscattering
probability for the deposit is much larger than for the substrate.

\begin{figure}
  \centering
  \includegraphics[width=0.5\textwidth]{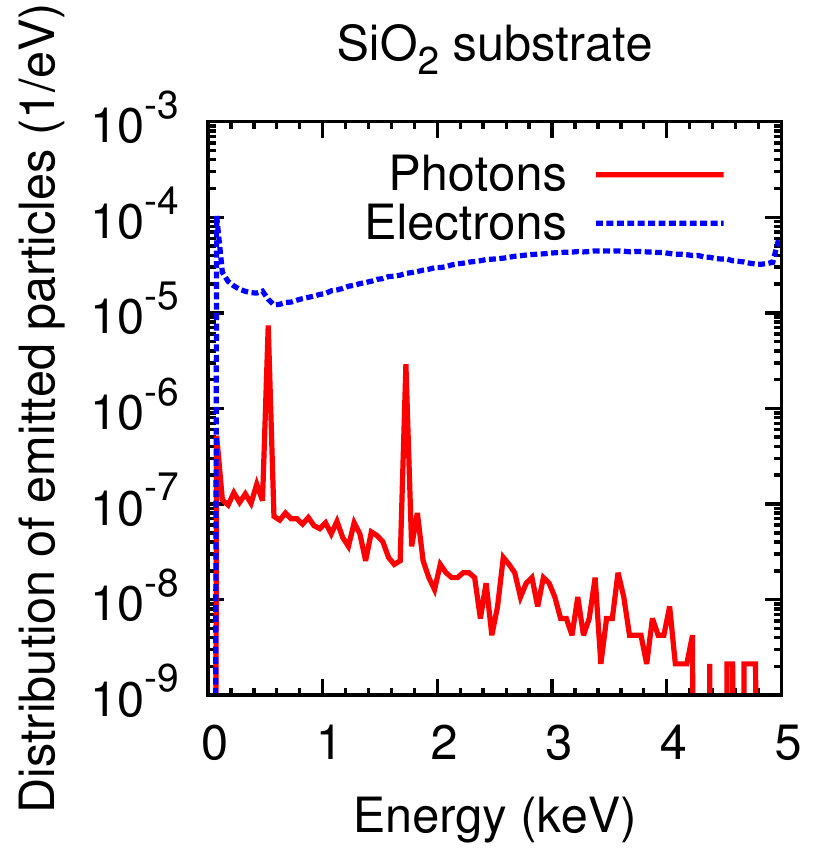}
  \caption{Distribution of electrons backscattered and emitted into the
    vacuum from the substrate in the absence of a deposit (dashed curve)
    and distribution of photons emitted into the vacuum (solid curve).
    Note, that the ordinates are in a logarithmic scale, whereas the
    abscissas are in a linear scale.}
  \label{fig:photons}
\end{figure}

The primary energy of the electrons (5 keV) is high enough to produce
inner-shell ionizations in Si and O. Let S0 denote the ionized shell.
A second electron from an outer shell, S1, fills the vacancy and,
subsequently, two processes are possible: (1) a radiative transition
whereby a photon is emitted with a characteristic energy
$U_\text{S0}-U_\text{S1}$, where $U$ denotes the ionization energy of
the corresponding shell, or (2), typically more likely, a non-radiative
transition whereby an electron from an outer shell S2 (which can either
coincide with or be less bound than S1) is emitted as an Auger electron
with energy $U_\text{S0}-U_\text{S1}-U_\text{S2}$. The emitted photons
might either leave the sample or be absorbed by a target atom, leading
to photoelectron emission. Figure~\ref{fig:photons} displays the
distribution of electrons (solid red curve) and photons (dashed blue
curve) emitted from the substrate per incoming electron in the absence
of a deposit. The peaks in the photon spectrum, superimposed on a
bremsstrahlung background, correspond to K lines of Si and O, situated
at 1739--1835~eV and 523~eV, respectively. Notice that the number of
photons emitted per incoming electron is at least two orders of
magnitude smaller than the number of emitted electrons. Furthermore,
interaction mean free paths for photons are typically much longer than
for electrons. Thus, the contribution of the emitted photons to the
energy and charge deposition processes is presumably negligible, except
for the minor photoelectron peak in the electron spectrum of the
substrate at 500 eV, superimposed to a contribution from Auger electron
emission from O with energies from 478.8 eV to 508.9 eV. A contribution
of Auger- or photoelectrons is not observed at $1739-1835$ eV, because
(1) Auger-electron energies are spread over a few hundred eV and thus do
not lead to a well resolved peak and (2) the photoelectric cross section
at these energies one order of magnitude smaller than at 500 eV (cross
section data taken from the numerical database of
\textsc{penelope}~\cite{salvat}). The photon spectrum was also
examined when a deposit lies on the substrate without finding
significant deviations regarding the minor role played by photon
transport as demonstrated for the pure substrate.

\begin{figure}
  \centering
  \includegraphics[width=0.9\textwidth]{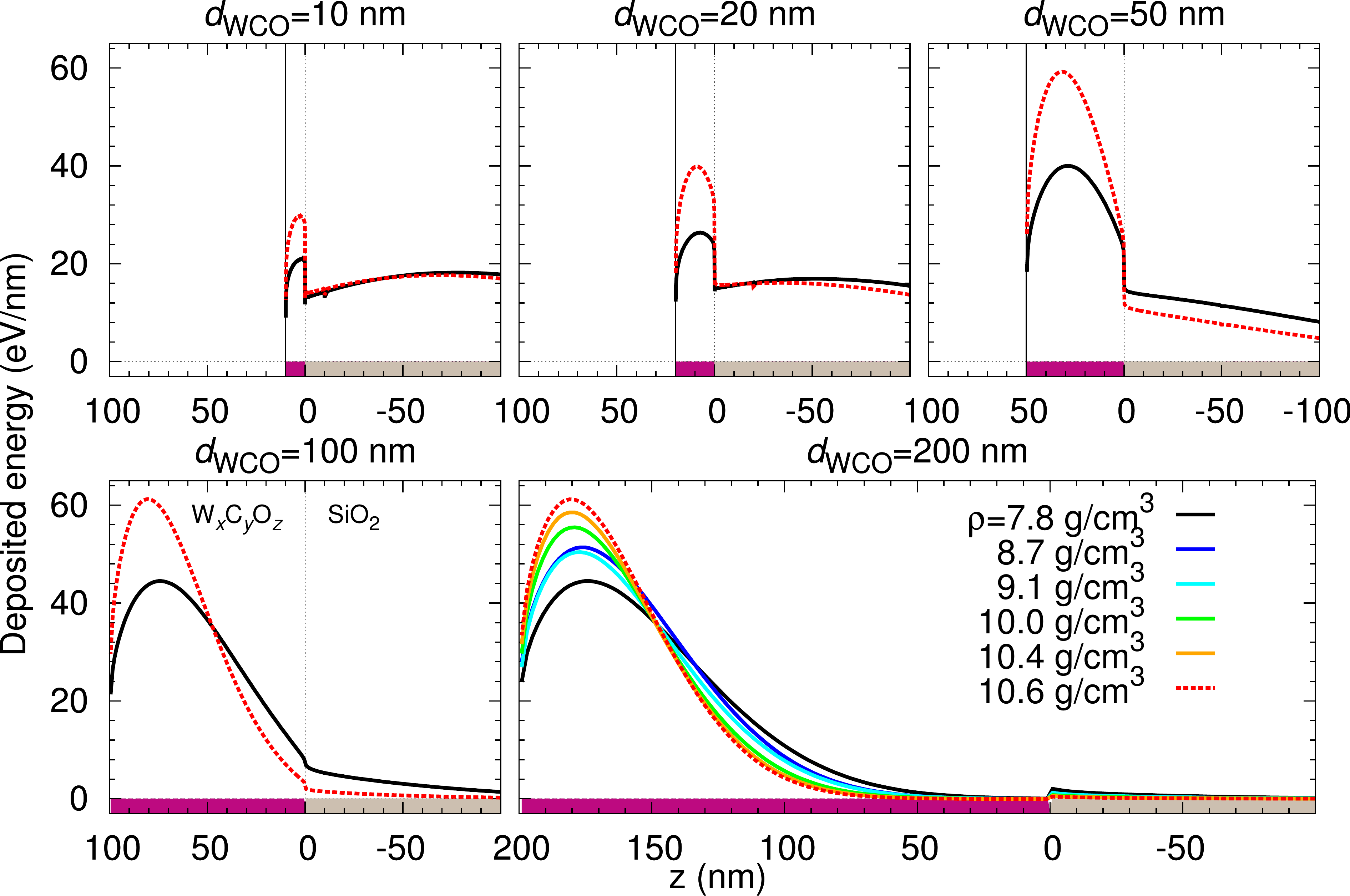}
  \caption{Energy deposited in the system as a function of the depth $z$
    for the indicated sample thicknesses $d_\text{WCO}$ and for the six
    nanostructure materials specified in Table~\ref{tab:dens}. The
    position $z=0$ corresponds to the deposit-substrate interface; the
    position $z=d_\text{WCO}$ (indicated by a solid vertical line in the
    three upper panels) corresponds to the deposit-vacuum interface.
  Notice that, in addition to the density, the composition of the
samples varies (see Table~\ref{tab:dens}).}
  \label{fig:dene}
\end{figure}

Figure~\ref{fig:dene} displays the distribution of energy deposited in
the system as a function of depth for sample thicknesses
$d_\text{WCO}$ ranging from 10~nm to 200~nm. Negative depths
correspond to the \ce{SiO$_2$} substrate, whereas positive depths
denote the deposit, indicated respectively by the magenta and grey bars
(reflecting the color code in Figure~\ref{fig:geometry}). The black
solid and the dashed red curve correspond to WC$_{2.5}$O (lowest
density sample) and WC$_{1.75}$O$_{0.75}$ (highest density sample),
respectively. The panel corresponding to $d_\text{WCO}=200$ nm
additionally shows the deposited energy for samples with intermediate
values of the density. It is clear that the deposited energy per unit
depth is much higher in the deposit than in the substrate, since the
IMFP is a factor $\sim2$ shorter in the deposit than in the substrate,
and thus energy-loss events take place more often in the deposit than
in the substrate. This also explains the discontinuous jump at the
deposit-substrate interface.

\begin{figure}
  \centering
  \includegraphics[width=0.9\textwidth]{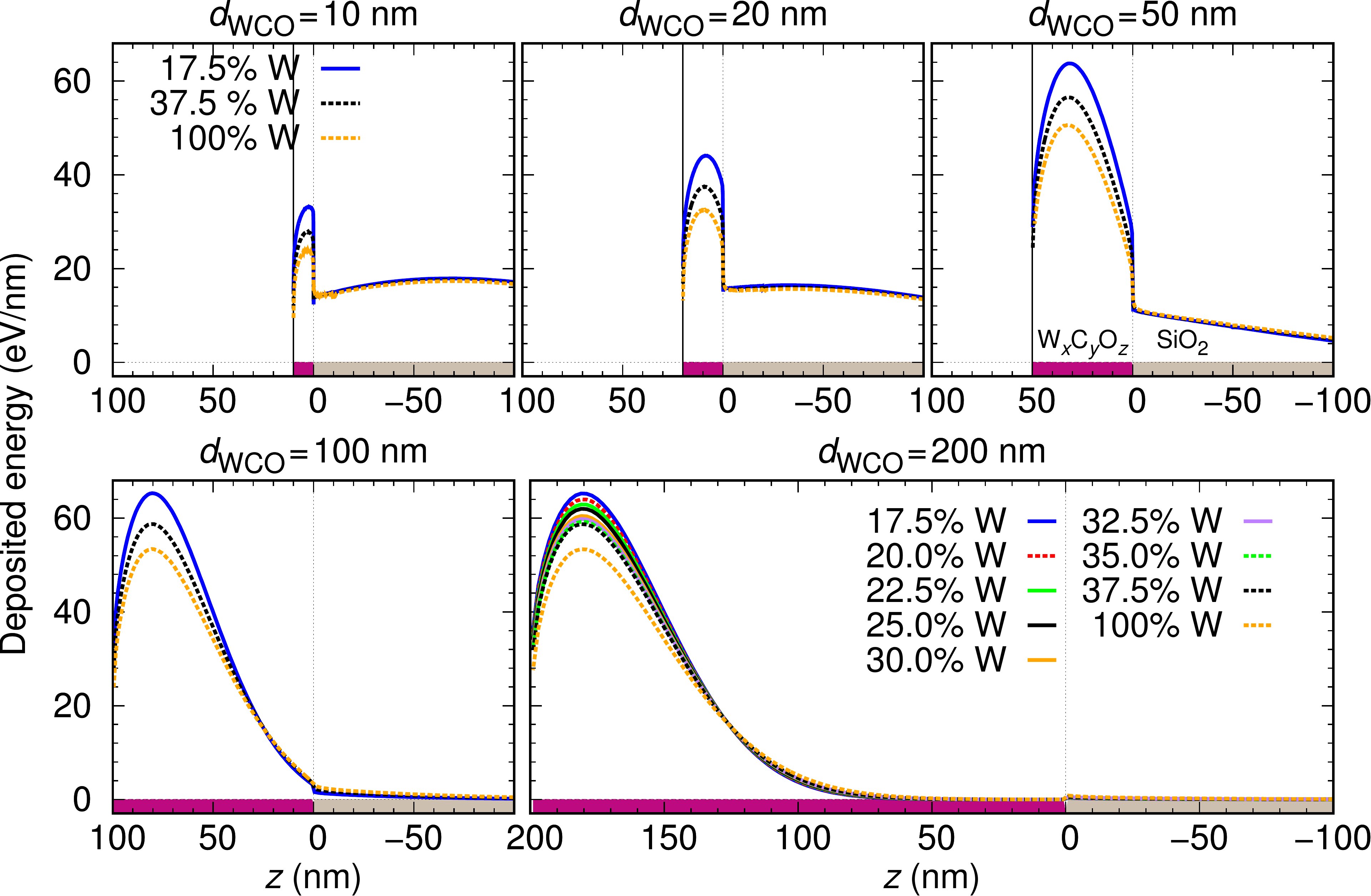}
  \caption{Same as Figure~\ref{fig:dene} for a fixed deposit density
  ($\rho=10.6092$ g/cm$^3$) and variable tungsten content.}
  \label{fig:denew}
\end{figure}

\begin{figure}
  \centering
  \includegraphics[width=0.5\textwidth]{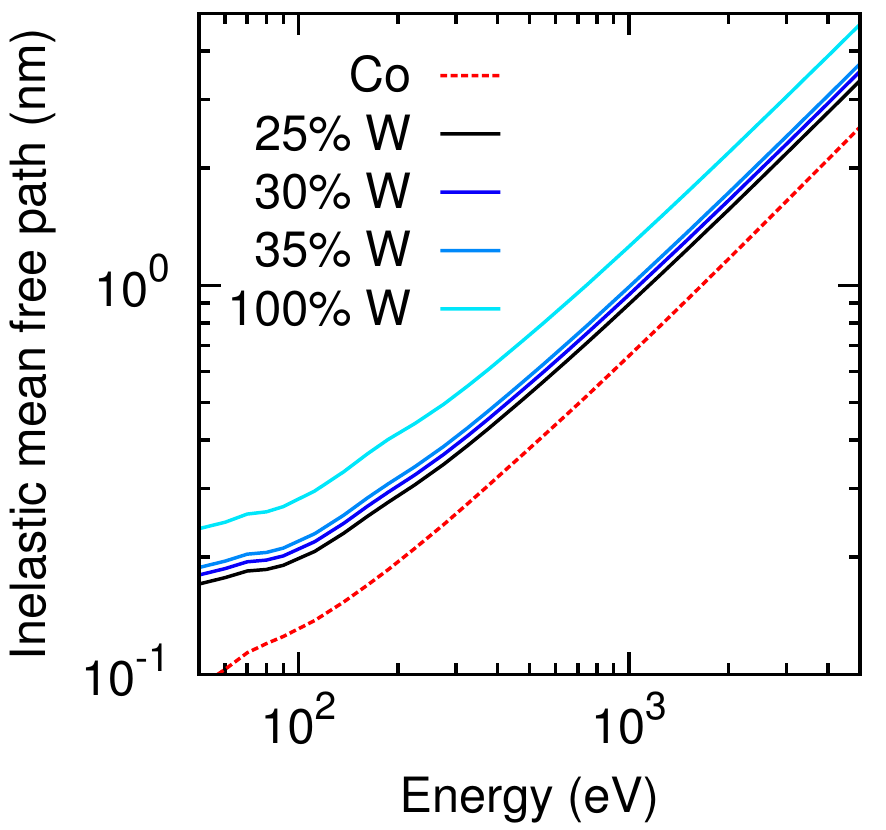}
  \caption{IMFP in the different deposits with fixed density
  ($\rho=10.6092$ g/cm$^3$) and variable tungsten content. }
  \label{fig:imfpw}
\end{figure}

\begin{figure}
  \centering
  \includegraphics[width=0.9\textwidth]{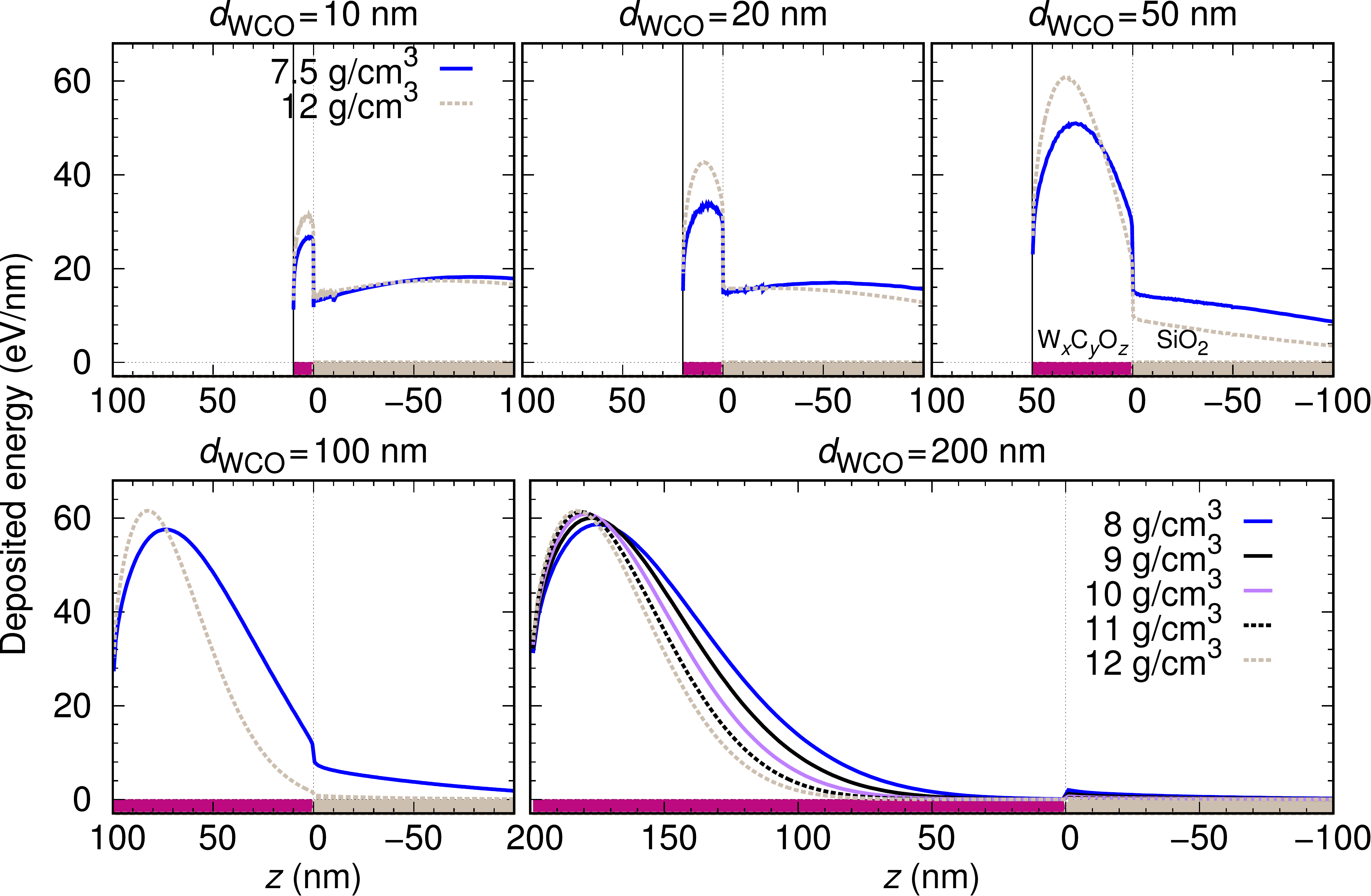}
  \caption{Same as Figure~\ref{fig:dene} for a fixed composition (27.5\%
    W, 50.4\% C, 22.1\% O) and variable density.}
  \label{fig:denew2}
\end{figure}

\begin{figure}
  \centering
  \includegraphics[width=0.5\textwidth]{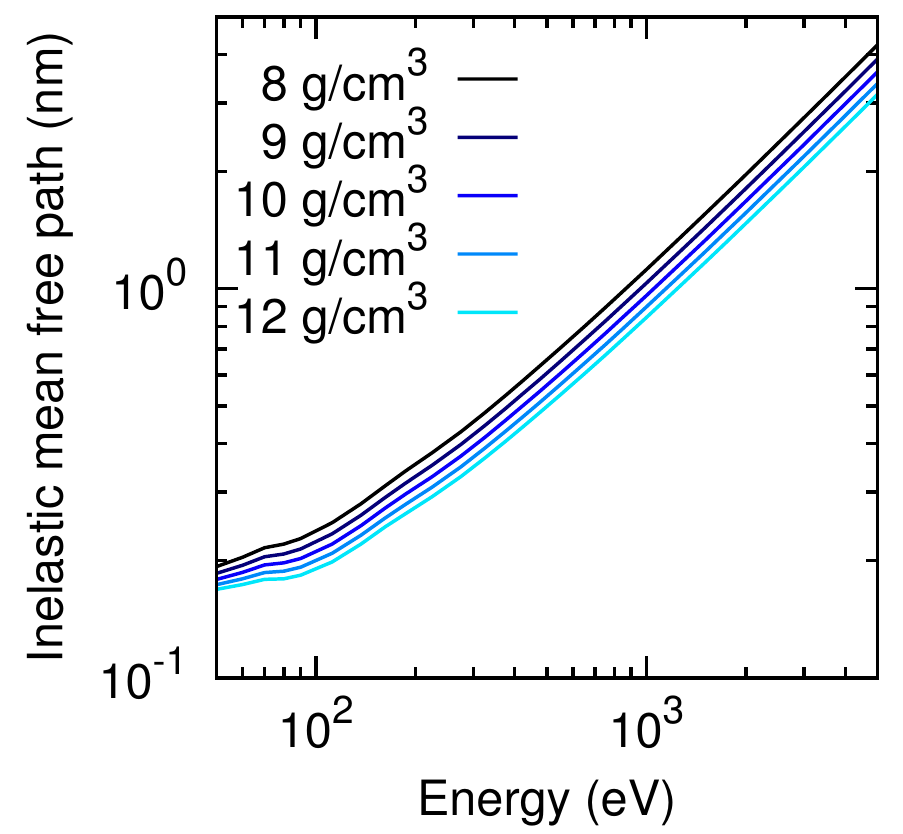}
  \caption{IMFP in different deposits with fixed composition (27.5\%
    W, 50.4\% C, 22.1\% O) and variable density.}
  \label{fig:imfpw2}
\end{figure}

It should be noted that, whereas the density increases linearly from
WC$_{2.5}$O to WC$_{1.75}$O$_{0.75}$, the tungsten content does not
exhibit a clear trend (see Table~\ref{tab:dens}). In order to
separately exhibit the effect of density and W-content variations on
the distribution of deposited energy, we have considered the following
artificial material variations. On the one hand, we have taken a sample with a fixed density
$\rho=10.6092$ g/cm$^3$ (corresponding to WC$_{1.75}$O$_{0.75}$) and
have varied its W content from 17.5\% to 37.5\% in steps of 2.5\%
(covering the range of W contents in Table~\ref{tab:dens}), decreasing
both the C and the O contents by 1.25\% at each step. We have also
considered the extreme case of 100\% W content. The distribution of
deposited energy as a function of depth is shown in
Figure~\ref{fig:denew} and the corresponding IMFPs are displayed in
Figure~\ref{fig:imfpw}. On the other hand, we have taken a sample with
a fixed W content (27.5\% W, 50.4\% C, 22.1\% O, corresponding to
WC$_{1.75}$O$_{0.75}$) and have varied its density from 8~g/cm$^3$ to
12~g/cm$^3$ (covering the range of densities given in
Table~\ref{tab:dens}). The distribution of deposited energy and the
corresponding IMFPs are shown in Figures~\ref{fig:denew2} and
\ref{fig:imfpw2}, respectively.  Comparing Figures~\ref{fig:denew} and
\ref{fig:denew2} we conclude that variations in the density influence
the energy deposition process in the nanostructure much more strongly
than variations in the W content (in the considered variation
intervals of these parameters). This can be best observed in the case
of sample thickness $d_\text{WCO} = 200$~nm for $z= 50$ to 150~nm.

\begin{figure}
  \centering
  \includegraphics[width=0.8\textwidth]{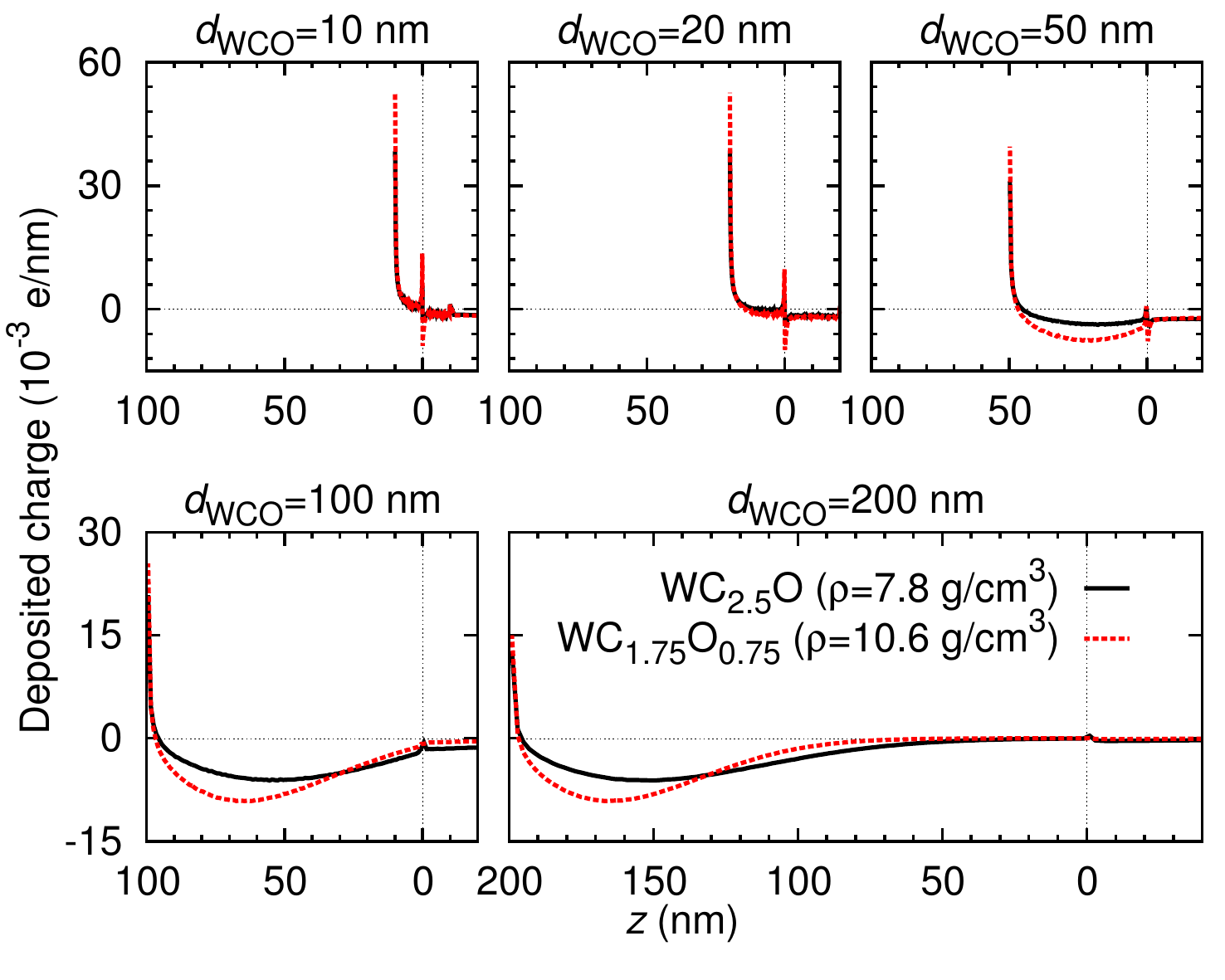}
  \caption{Charge deposited into the system as a function of the depth
    $z$ for the indicated sample thicknesses $d_\text{WCO}$ and for
    the six nanostructure materials specified in Table~\ref{tab:dens}
    (compare with Figure~\ref{fig:dene}).}
  \label{fig:dcharge}
\end{figure}

In practice, sample charging effects in the EBID process cause only a
minor repulsion of the electron beam (observed as a slight drift in the
monitoring images) which can be easily corrected by applying
appropriate beam-deflection voltages. Nevertheless it is interesting
\textit{per se} to examine the spatial distribution of the charge
deposition process induced by the incoming beam, if only to better delimit
the spatial region that is probed and affected by the
beam. Figure~\ref{fig:dcharge} displays the distribution of charge
deposited per unit path length for deposit thicknesses $d_\text{WCO}$
ranging from 10~nm to 200~nm in WC$_{2.5}$O and WC$_{1.75}$O$_{0.75}$
(solid black and dashed red line, respectively). Calculations were
also carried out for samples of intermediate densities but are
not shown in the Figure, which displays only the two extreme cases for
clarity. Note, that the charge deposited in the nanostructure close to
the vacuum interface is positive. This implies that there are more
secondary electrons emitted from this region than slow electrons
absorbed in it. Indeed, those secondary electrons emitted from the
nanostructure into the vacuum do not return, implying that close to
the vacuum interface it is more likely to see a lack of electrons than
the absorption of slow electrons.  Deeper into the nanostructure, the
absorption of slow electrons becomes more likely: secondary electrons
are more likely to become absorbed than to reach the interface into
vacuum. This leads to the observed decrease in the deposited charge,
which becomes even negative when the nanostructure is thick enough so
that all generated secondary electrons are eventually absorbed in
it. Regarding the deposit-substrate interface, two aspects should be
considered. On the one hand, the primary electron loses less energy in
the SiO$_2$ substrate than in the nanostructure material, so that less
secondary electrons are generated per unit path length in the
substrate.  On the other hand, secondary electrons from the
nanostructure cross the interface into the substrate. The number of
slow electrons moving from the nanostructure into the substrate is
larger than the number of slow electrons moving in the opposite
direction. This leads to the observed increase in positive (negative)
charge in the nanostructure (substrate) side of the interface.

\begin{figure}
  \centering
  \includegraphics[width=0.9\textwidth]{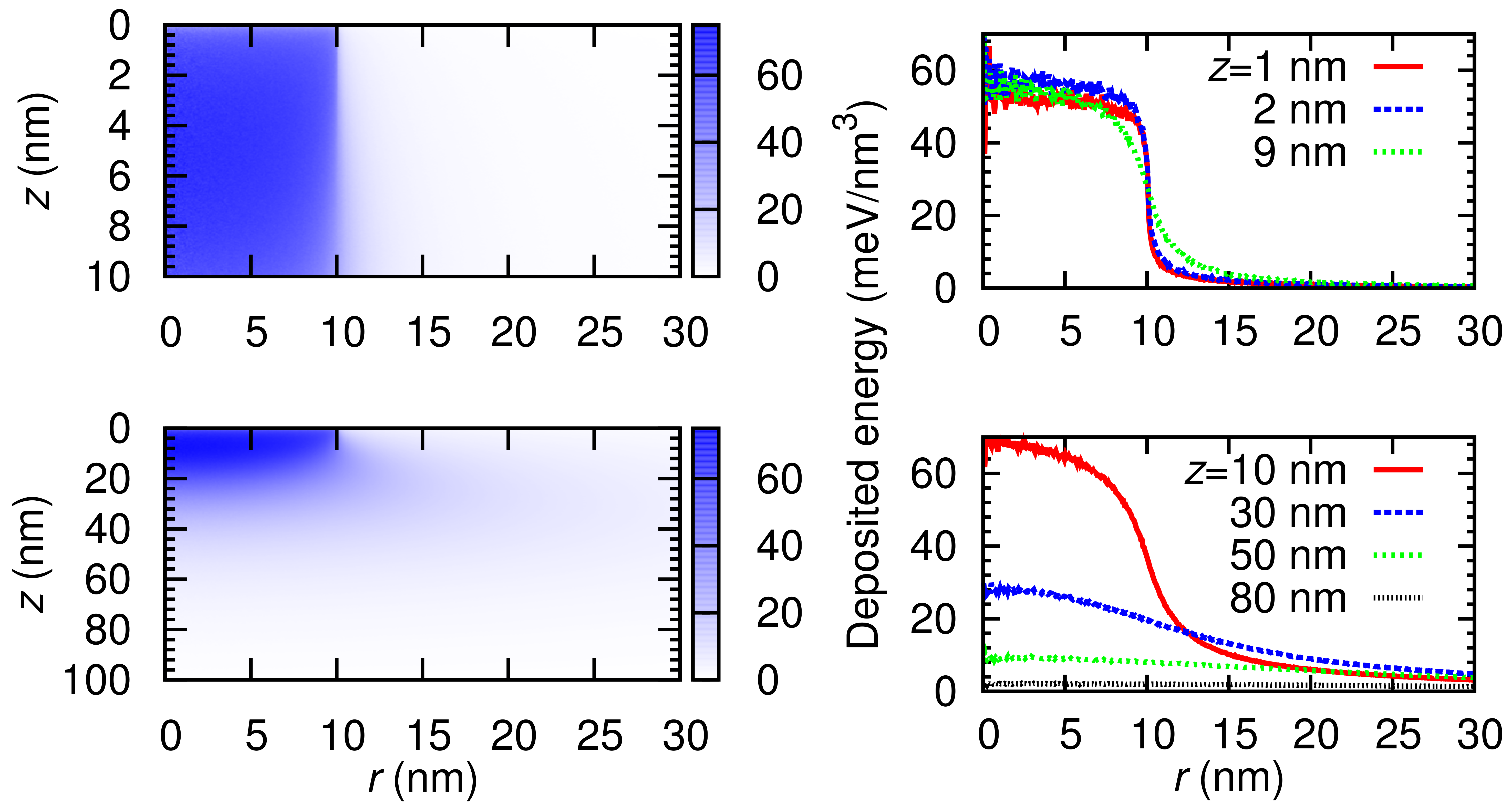}
  \caption{Distribution of energy deposited in WC$_{1.75}$O$_{0.75}$ as
    a function of the depth $z$ and the radial coordinate $r$. The two
    upper (lower) panels correspond to a sample thickness
    $d_\text{WCO}=10$~nm ($d_\text{WCO}=100$~nm). The right-hand-side
    panels display cross sections of the distribution at the indicated
    depths $z$ below the deposit-vacuum interface.}
  \label{fig:2de}
\end{figure}

\begin{figure}
  \centering
  \includegraphics[width=0.9\textwidth]{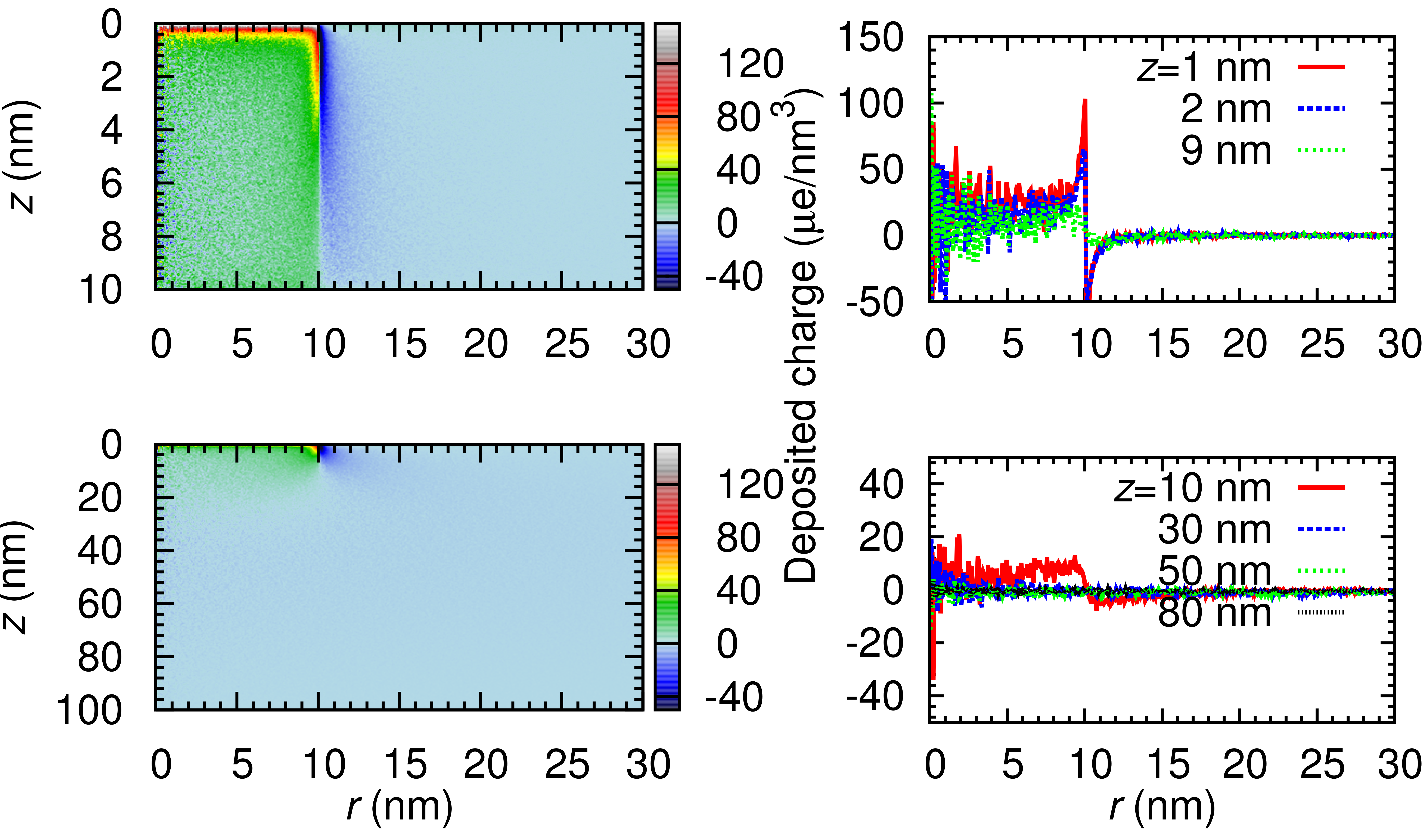}
  \caption{Same as Figure~\ref{fig:2de} for the charge deposited in
  WC$_{1.75}$O$_{0.75}$.}
  \label{fig:2dc}
\end{figure}

Finally, Figures~\ref{fig:2de} and \ref{fig:2dc} display, respectively,
the distribution of deposited energy and charge as a function of the
depth and the radial coordinate in WC$_{1.75}$O$_{0.75}$ (highest
density sample) for nanostructure thicknesses of 10 and 100 nm. The
panels in the right-hand side show cross sections of the distributions
at the indicated depths $z$. In these figures one can clearly identify
the beam radius of 10 nm. Notice that at radii $r<10$ nm the deposited
charge is positive. In this region, secondary electrons are emitted as
a result of the energy loss of the primary electrons. For distances
$r>10$ nm, the deposited charge is negative, meaning that electrons
with $E\leq 50$ eV are absorbed there. These slow electrons are those
secondary electrons generated in $r<10$ nm which wander into $r>10$ nm
and are not able to travel further, being absorbed.

The distribution of deposited energy as a function of the depth and of
the radial coordinate has additional value. On the one hand, it can be
used to derive a temperature distribution for more detailed microscopic
simulations (\textit{e.g.} molecular dynamics) of the EBID process
\cite{randolph2005}. On the other hand, the deposited energy also
contributes to an enhancement of the dissociation of precursor gas
molecules adsorbed on the surface. It is therefore worthwhile to have
an accurate estimate for this quantity. The consequences for
nanostructure growth are an interesting topic of further research. 

\section{Conclusions}

In this work we presented results of Monte Carlo simulations of
electron transport which provide valuable insight into the charge and
energy deposition processes induced by the primary electron beam in
the EBID process of W(CO)$_6$ nanostructures on SiO$_2$
substrates. The simulations highlight the differences in the transport
of electrons in the nanostructure and in the substrate: the mean free
path between consecutive inelastic interactions in the deposit is a
factor $\sim2$ smaller than in the substrate which leads to a beam
attenuation after a depth of $\sim500$~nm in the substrate material,
whereas, in the nanostructure material, the beam is attenuated at much
shallower depths of $\sim150$~nm.  In the early stages of the
nanostructure growth (thickness well below 150~nm), a significant
fraction of incoming electron trajectories still interact with the
substrate. As the nanostructure becomes thicker ($\gtrsim100$ nm), the
transport takes place almost exclusively in the nanostructure, leading
to a saturated behavior of the distribution of the deposited energy,
charge, and backscattered electrons. The simulations show two effects
which may be important for the growth of the nanostructure.
(1) The energy deposited in the substrate is available for the
dissociation of precursor-gas molecules adsorbed on the surface
substrate. (2) If we assume that a larger yield of secondary electrons
enhances precursor dissociation and improves the conditions for
nanodeposit growth with high density and metal content, then the
simulations show that larger deposit density leads to enhanced electron
backscattering. This implies that random fluctuations in deposit density
could be amplified through positive feedback.

The presented simulations therefore provide an overview of the effect of
the primary-electron beam on the deposit and on the substrate at
different stages of the nanostructure growth. Furthermore, the
distributions of deposited energy serve as a starting point for further
microscopic simulations (molecular dynamics) in that they provide a
guideline for the initial temperature distribution in the substrate and
the deposit under irradiation with an electron beam. Moreover, similar
simulations can aid in understanding the role played by backscattered
and secondary electrons in changing structural properties of
nanostructured materials in several post-growth techniques,
including direct or oxygen-assisted electron-beam curing
\cite{porrati2011,mehendale2013}.

\acknowledgements We are indebted to Prof.\ Francesc Salvat of the
Universitat de Barcelona for the very fruitful discussions and we
acknowledge the contribution of Carlos Ortiz at the beginning of this
work. Funding by the Beilstein-Institut is gratefully acknowledged.

\bibliography{refs}
\vspace{3cm}

\end{document}